\documentstyle[pra,preprint,aps]{revtex}

\begin{document}
\title{Observation of sub-natural linewidths for cold atoms in a 
magneto-optic trap}
\author{Umakant D. Rapol, Ajay Wasan, and Vasant 
Natarajan\thanks{E-mail: vasant@physics.iisc.ernet.in}}
\address{Department of Physics, Indian Institute of Science, 
Bangalore 560 012, INDIA}

\maketitle

\begin{abstract}
We have studied the absorption of a weak probe beam through cold 
rubidium atoms in a magneto-optic trap. The absorption spectrum 
shows two peaks with the smaller peak having linewidth as small as 
28\% of the natural linewidth. The modification happens because 
the laser beams used for trapping also drive the atoms coherently 
between the ground and excited states. This creates ``dressed'' 
states whose energies are shifted depending on the strength of the 
drive. Linewidth narrowing occurs due to quantum coherence between 
the dressed states. The separation of the states increases with 
laser intensity and detuning, as expected from this model.
\end{abstract}

The use of near-resonant laser light to cool atoms from room 
temperature to ultra-low temperatures has become a standard 
technique for atomic physics experiments \cite{nobel97}. 
Temperatures in the range of few microkelvin are routinely 
achieved using these methods. The cooled atoms can also be trapped 
by superposing a magnetic field on the laser beams in a 
configuration called a magneto-optic trap (MOT) \cite{RPC87}. Cold 
atoms in a MOT form near-ideal laboratories to test our 
understanding of the fundamental laws of physics. There are two 
primary reasons for this. The first is that almost all laser 
cooling experiments are done on alkali atoms, and alkali atoms are 
attractive candidates for such fundamental experiments because 
their simple one-electron structure renders them amenable to 
theoretical calculations. For example, the alkali atom cesium has 
been used for high-precision experiments on atomic parity 
violation \cite{WBC97} because the experimental results can be 
compared to theoretical predictions based on the Standard Model. 
The second advantage of cold atoms in a MOT is that, if one is 
looking for spectroscopic signatures of some effect, then the use 
of ultra-cold atoms guarantees high precision because Doppler 
broadening is negligible. For example, a probe beam sent through a 
cloud of rubidium atoms in a MOT shows a Doppler width of 500 kHz, 
while the natural linewidth of the cooling transition is 6.1 MHz. 
However, the natural linewidth appears as a fundamental limit to 
the resolution achievable in these experiments. In addition, it is 
often necessary to turn off the MOT before doing any precision 
measurement because the trapping beams cause unwanted perturbation 
to the energy levels of the atoms. In this article, we show that, 
not only can the energy shift from the trapping beams be well 
understood from a simple theoretical model, but this can also lead 
to significant narrowing of the linewidth of the transition below 
the natural linewidth. 

The MOT was first invented in 1987 \cite{RPC87} and since then it 
has become the workhorse for any experiment that requires a cold 
and dense cloud of atoms. It is the starting point for experiments 
ranging from atomic fountain clocks to achieving Bose-Einstein 
condensation \cite{AEM95,KET99}. We have been interested in a MOT 
primarily from its potential use in precision spectroscopy 
experiments. The standard MOT configuration consists of three 
pairs of counter-propagating laser beams in the three orthogonal 
directions, and a quadrupole magnetic field generated using two 
coils carrying current in opposite directions. The laser beams are 
circularly polarized and detuned below the resonance by 2 to 
$3\Gamma$, where $\Gamma$ is the natural linewidth of the cooling 
transition.

We work with $^{87}$Rb atoms in the MOT, for which the cooling 
transition is the $5S_{1/2},F=2 \leftrightarrow 5P_{3/2},F'=3$ 
transition at 780 nm. The transition has a natural linewidth of 
6.1 MHz and saturation intensity \cite{foot1} of 1.6 mW/cm$^2$. 
The laser beams are derived from a frequency-stabilized diode 
laser system locked near this transition. To prevent the atoms 
from optically pumping out of the cooling transition, a small 
amount of ``repumping'' light resonant with the $5S_{1/2},F=1 
\leftrightarrow 5P_{3/2},F'=2$ transition is mixed with the 
primary MOT beams. The repumping beam is obtained from a second 
tunable diode laser system. The experiments are done in a vacuum 
chamber maintained at a pressure below $10^{-9}$ torr. Rb atoms 
are loaded into the trap using a getter source \cite{RWN01} that 
fills the chamber with Rb vapor when heated with a few amps of 
current. The low-energy tail from this background vapor is cooled 
and captured in the MOT.

The loading of atoms into the MOT follows an exponential growth 
curve \cite{RWN01}. The time constant is a few seconds, so that, 
after about a minute, the MOT is in steady state with a number of 
atoms that stays constant as long as the source remains on. Under 
typical conditions, the steady-state MOT has about $10^8$ atoms at 
a temperature of 300 $\mu$K and density of $10^{10}$ atoms/cc. We 
have studied the absorption of a weak probe beam passing through 
the cold atoms in a MOT after it has attained steady state. The 
probe is linearly polarized and has an intensity of about 250 
$\mu$W/cm$^2$. A typical absorption spectrum as the probe laser is 
scanned across the $5S_{1/2},F=2 \leftrightarrow 5P_{3/2},F'=3$ 
transition is shown in Fig.\ \ref{f1}. The main features of the 
spectrum are that there are two peaks instead of one, and that the 
linewidth of the smaller peak is only 1.7 MHz, or about 28\% of 
the natural linewidth.

To understand this we have to consider that, in the presence of 
the probe and trapping beams, the Rb atoms form an effective 
three-level system as shown in Fig.\ \ref{f2}. In other words, the 
circularly-polarized trapping beam couples certain magnetic 
sublevels in the ground and excited state, while the 
linearly-polarized probe beam measures absorption from the same 
ground-state sublevel to a different excited-state sublevel. For the 
trapping beam, we only consider the stretched state, {\it i.e.\/} 
the magnetic sublevel where the projection of angular momentum 
($m_F$) is maximum. The main reason for this is that, as the atoms 
move away from the trap center, they rapidly get optically pumped 
into this state. Once they are in this state they remain there 
until they go to the other side of the trap where they get 
optically pumped into the opposite stretched state. For example, 
the $\sigma^+$ beam optically pumps atoms into the 
$5S_{1/2},F=2,m_F=2$ sublevel from where they cycle on the 
$F=2,m_F=2 \leftrightarrow F'=3,m_{F'}=3$ transition, while on the 
other side of the trap, the $\sigma^-$ beam optically pumps atoms 
into the $5S_{1/2},F=2,m_F=-2$ sublevel. 

The primary purpose of the trapping beam is to provide a radiation 
pressure force that pushes the atoms towards the trap center. This 
force arises because the atom absorbs a laser photon, goes to the 
excited state, and then decays to the ground state through {\it 
spontaneous emission}. However, there is a small probability that 
the atom decays through {\it stimulated emission} induced by the 
same beam. The resulting force is called the stimulated force and 
causes the atom to be coherently driven between the ground and 
excited states. The frequency at which the probability swaps 
between the two states is called the Rabi frequency, $\Omega_R$, 
and is determined by the laser intensity and the strength of the 
transition \cite{foot2}. Since we are going to be interested only 
in this coherent driving of the atom, this gives another 
justification for considering only the stretched state in our 
analysis: transitions starting from this state are the strongest 
in terms of relative oscillator strength. 

The energy levels shown in Fig.\ \ref{f2} can now be understood 
more clearly. The $\sigma^+$ trapping beam couples the 
$5S_{1/2},F=2,m_F=2$ and $5P_{3/2},F'=3,m_{F'}=3$ levels, while 
the probe measures absorption on the $5S_{1/2},F=2,m_F=2 
\rightarrow 5P_{3/2},F'=3,m_{F'}=2$ transition. The trapping beam 
has a detuning from resonance of $\Delta_t$ and the probe beam has 
a detuning of $\Delta$. The spontaneous decay rates from the 
excited levels, $\Gamma_{21}$ and $\Gamma_{31}$, are both equal to 
$\Gamma$, which is 6.1 MHz. As discussed above, the trapping beam 
partly drives the atoms coherently. It is well known that this 
driving creates ``dressed'' states whose energies are shifted by 
the strength of the drive (also called the ac Stark shift) 
\cite{COR77}. The energy shift is analogous to the shift in the 
frequency of two classical oscillators that are coupled together, 
where the shift is proportional to the strength of the coupling. 
In fact, this analogy with coupled classical oscillators can be 
made more exact, and results in analogous definitions of classical 
dressed states and Rabi frequency \cite{CWB90}.

In the presence of the dressed states, the probe-absorption 
spectrum splits into two peaks, called an Autler-Townes doublet 
\cite{AUT55}. The location of the two peaks is given by 
\cite{AGA96,VAR96}:
\begin{equation}
\label{e1}
\Delta_{\pm} = \frac{\Delta_t}{2} \pm 
\frac{1}{2}\sqrt{\Delta_t^2 + \Omega_R^2},
\end{equation}
where $\Delta_{+}$ and $\Delta_{-}$ are the values of the probe 
detuning where the peaks occur. The corresponding linewidths 
($\Gamma_{\pm}$) of these peaks are different because of the 
coherence between the two dressed states, and given by
\begin{equation}
\label{e2}
\Gamma_{\pm} = \frac{\Gamma}{2} 
\left( 1 \mp \frac{\Delta_t}{\sqrt{\Delta_t^2 + \Omega_R^2}} 
\right) .
\end{equation}
It is clear from the above expression that, if $\Delta_t = 0$, the 
two peaks are symmetric and have identical linewidths of 
$\Gamma/2$. However, for any non-zero detuning, the peaks have 
asymmetric linewidths. The first peak has larger linewidth while 
the second peak has smaller linewidth by precisely the same 
factor, in such a way that the sum of the two linewidths is equal 
to the unperturbed linewidth, $\Gamma$.

The above analysis is for a stationary atom. If the atom is 
moving, the laser detuning as seen by the atom depends on its 
velocity. To obtain the probe absorption in a gas of moving atoms, 
the above expressions have to be corrected for the velocity of the 
atom and then averaged over the Maxwell-Boltzmann distribution of 
velocities. Such an analysis has been done by Vemuri {\it et 
al.\/} in Ref.\ \cite{VAR96}. The important conclusion of that 
work is that the location of the peaks given in Eq.\ \ref{e1} does 
not change, but the linewidths are now given by
\begin{equation}
\label{e3}
\Gamma_{\pm} = \frac{\Gamma + D}{2} 
\left( 1 \mp \frac{\Delta_t}{\sqrt{\Delta_t^2 + \Omega_R^2}} 
\right) . 
\end{equation}
Here, $D$ is the Doppler width, which is 0.5 MHz for Rb atoms at a 
temperature of 300 $\mu$K. This is negligible compared to the 
natural linewidth of 6.1 MHz. Therefore, for our conditions, Eqs.\ 
\ref{e1} and \ref{e2} can be taken as valid even at the finite 
temperature in the trap.

The probe-absorption spectrum presented in Fig.\ \ref{f1} can now 
be understood based on the above analysis. The trapping beam 
coherently drives the atoms and creates two dressed states near 
the original ground state. The probe beam measures absorption from 
the modified ground state and therefore shows two peaks. For 
non-zero detuning of the trapping beam, the linewidth of the smaller 
peak is much smaller than $\Gamma$, as given in Eq.\ \ref{e2}. 

In order to verify that this model is correct, we have studied the 
probe absorption at different values of trapping-beam intensity 
(or Rabi frequency) and detuning. From Eq.\ \ref{e1}, the 
separation of the two peaks in the absorption spectrum is 
$\sqrt{\Delta_t^2 + \Omega_R^2}$, and should increase when either 
$\Omega_R$ or $\Delta_t$ is increased. The variation with laser 
intensity at three different values of $\Delta_t$ is shown in 
Fig.\ \ref{f3}. The solid line is a fit to the variation expected 
from Eq.\ \ref{e1} and matches the data quite well. However, for 
each curve, the $y$-intercept should be the detuning $\Delta_t$. 
Instead, we find that the intercept is smaller by about 4 MHz. 
There could be several reasons for this. One likely explanation is 
that, because the atoms are in a linearly-varying magnetic field, 
the detuning of the trapping laser depends on the exact location 
of the atom. Atoms that get optically pumped into the stretched 
state are most likely to be near the periphery of the cloud. 
Taking into account the 4 mm size of the cloud and the 
magnetic-field gradient of 10 G/cm, the field at this location is about 2 
G. For the stretched state, this would cause a decrease in 
detuning by 2.8 MHz, which is about the size of the decrease we 
measure. In addition, atoms away from the center of the trap have 
a non-zero velocity. Because the lasers are detuned below 
resonance, optical pumping into the stretched state is most likely 
when the atoms are moving towards the laser beam. The Doppler 
shift in this situation is such that it would again cause a 
reduction in the detuning. For atoms moving with the r.m.s.\ 
velocity in the trap, the Doppler shift is about 0.4 MHz. 

Whatever be the exact cause for this change in detuning seen by 
the atoms, we can verify its effect further by studying the 
separation of the peaks as a function of detuning. In Fig.\ 
\ref{f4}, we show the separation {\it vs.\/} detuning for 
different values of trapping-beam intensity. The solid lines are 
exact calculations from Eq.\ \ref{e1} and the measured points lie 
almost perfectly on these curves. However, to get this match, we 
had to use the effective detuning from the $y$-intercepts of Fig.\ 
\ref{f3}, and not the zero-field detuning set in the laboratory. 
The excellent agreement with the theoretical prediction gives us 
confidence that the atoms are indeed seeing a smaller detuning.

There are other features of the absorption spectra that are 
clearly noticeable in the data. For example, the smaller peak 
becomes more prominent (relative to the larger peak) as the 
trapping-beam intensity is increased at a given detuning. 
Similarly, for a fixed intensity, the peak becomes more prominent 
at smaller detunings. Both these observations are consistent with 
the model of a driven two-level system, since the effect of the 
drive is stronger when the intensity is higher or when it is 
closer to resonance. However, there are some complications with 
this simple model that are also evident in the data. One problem 
is that the linearly-polarized probe laser can also couple to the 
$F'=3,m_{F'}=1$ sublevel, in addition to the $F'=3,m_{F'}=2$ 
sublevel already considered. The location of the $F'=3,m_{F'}=1$ 
absorption peak is slightly shifted due to the non-zero magnetic 
field. This transition is very weak in terms of relative strength 
compared to the transition to the $F'=3,m_{F'}=2$ level, therefore 
it would show up only in the primary absorption peak. But this 
implies that the primary peak is a convolution of two peaks. We 
see some evidence for this in the data because we find that the 
lineshape of the primary peak is non-Lorentzian. This makes it 
difficult to extract a reliable estimate of the linewidth of the 
peak but does not affect the location of the peak center that has 
been used for Figs.\ \ref{f3} and \ref{f4}. 

In conclusion, we have observed sub-natural linewidth for probe 
absorption through cold rubidium atoms in a MOT. The modification 
to the absorption spectrum arises because of dressed states 
created by the trapping beams. Perhaps the most useful way to 
understand this is in the context of driven three-level systems 
\cite{NSO90,AGA96} and the phenomenon of electromagnetically 
induced transparency (EIT) \cite{HAR97}. In EIT, an initially 
absorbing medium is made transparent to the probe when a strong 
control laser is applied to an auxiliary transition. In our case, 
the trapping laser itself plays the role of the control laser that 
modifies the absorption of the probe. It thus opens up new 
possibilities for EIT experiments using cold atoms, where 
phenomena such as gain without inversion, anomalous dispersion, 
and population trapping can be studied.

\acknowledgments
This work was supported by a research grant from the Department of 
Science and Technology, Government of India.

\begin{figure}
\caption{
The figure shows the absorption of the probe beam as a function of 
its detuning from the $5S_{1/2},F=2 \leftrightarrow 5P_{3/2},F'=3$ 
transition. The important feature is the appearance of two peaks 
instead of one. The dotted line is a Lorentzian fit to the two 
peaks and yields linewidths (FWHM) of 14 MHz for the larger peak and only 
1.7 MHz ($0.28\Gamma$) for the smaller peak. The trapping beam 
detuning $\Delta_t$ is $-19$ MHz, and the intensity $I$ is 4.8 
mW/cm$^2$.
}
\label{f1}
\end{figure}

\begin{figure}
\caption{
Effective three-level system in $^{87}$Rb when both trapping and 
probe beams are on. The trapping laser is circularly polarized and 
optically pumps the atoms into the stretched state. For 
$\sigma^+$-pol, the trapping laser couples the $F=2,m_F=2$ and 
$F'=3,m_{F'}=3$ levels at a detuning $\Delta_t$. The probe laser 
is linearly polarized and measures absorption on the $F=2,m_F=2 
\rightarrow F'=3,m_{F'}=2$ transition at a detuning $\Delta$. 
$\Gamma_{31}$ and $\Gamma_{21}$ are the spontaneous decay rates.
}
\label{f2}
\end{figure}

\begin{figure}
\caption{
The figure shows the separation of the two absorption peaks as a 
function of trapping-beam intensity, for three values of detuning. 
The solid lines are fits to Eq.\ 1 in the text. From Eq.\ 1, the 
$y$-intercept in each case must be $\Delta_t$, but the value from 
the fit is lower by about 4 MHz. This indicates that the effective 
detuning seen by the atoms is smaller. Otherwise, the separation 
follows the predicted variation very closely. 
}
\label{f3}
\end{figure}

\begin{figure}
\caption{
The figure shows the separation of the two absorption peaks as a 
function of detuning, for three values of trapping-beam intensity. 
The solid lines are solutions to Eq.\ 1 in the text with no 
adjustable parameters, provided we use the effective detuning 
values obtained from the $y$-intercepts in Fig.\ 3 and not the 
values we set. Note the excellent agreement with theory under this 
assumption.
}
\label{f4}
\end{figure}


\begin{references}
\bibitem{foot1}
The saturation intensity is a measure of the strength of the 
laser-induced coupling between the ground and excited states. It 
is defined as that intensity for which the transition is power 
broadened by a factor of $\sqrt{2}$.
\bibitem{foot2}
The relation between the intensity and the Rabi frequency is 
$I/I_0 = 2\Omega_R^2 / \Gamma^2$, where $I_0$ is the saturation 
intensity.
\end{references}
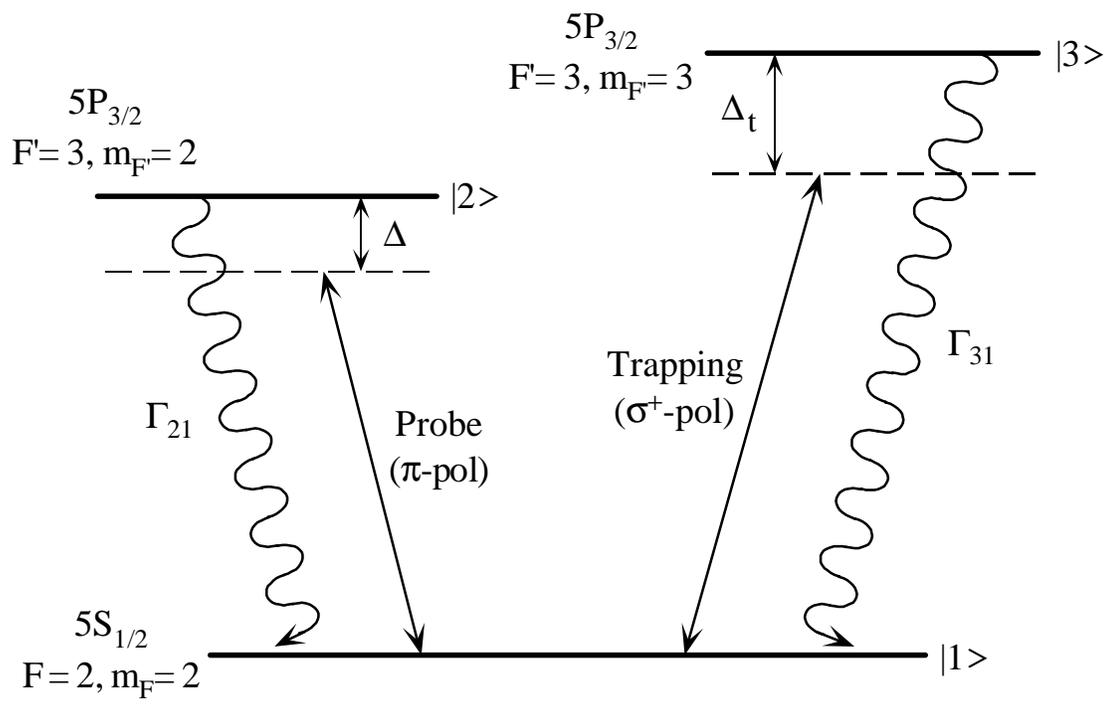
\end{document}